\documentclass[%
  reprint,            
  superscriptaddress, 
  amsmath,amssymb,    
  aps,                
  prl,                
  floatfix,           
  longbibliography    
]{revtex4-2}

\usepackage{graphicx}  
\usepackage{bm}        
\usepackage{dcolumn}   

\usepackage[colorlinks=true,
            linkcolor=blue,
            citecolor=blue,
            urlcolor=blue]{hyperref}

\usepackage{microtype} 
\usepackage{comment}
\usepackage[normalem]{ulem}
\newcommand{\ve}{\varepsilon}
\newcommand{\vf}{v_\mathrm{F}}
\newcommand{\nf}{\nu_\mathrm{F}}
\newcommand{\see}{\sigma_\mathrm{ee}}

\newcommand{\tee}{\tau_\mathrm{ee}}

\newcommand{\ti}{\tau_\mathrm{i}}
\newcommand{\ttot}{\tau_\mathrm{tot}}

\newcommand{\bk}{\mathbf{k}}
\newcommand{\bv}{\boldsymbol{v}}

\newcommand{\bq}{\mathbf{q}}

\newcommand{\bE}{\mathbf{E}}

\newcommand{\kf}{k_{\rm F}}
\newcommand{\ef}{\varepsilon_{\rm F}}
\newcommand{\beq}{\begin{equation}}
\newcommand{\eeq}{\end{equation}}
\newcommand{\bea}{\begin{eqnarray}}
\newcommand{\eea}{\end{eqnarray}}
\newcommand{\nn}{\nonumber}
\newcommand{\R}{\mathrm{Re}\,}

\newcommand{\la}{\left\langle}
\newcommand{\ra}{\right\rangle}

\newenvironment{centeq}
    {\begin{equation} \begin{aligned}}
    {\end{aligned} \end{equation}}

\makeatletter
\begin{document}
\title{Conserving relaxation-time approximation for electron-electron collisions}
\author{Sayak Bhattacharjee}
\email{sayakbhattacharjee@stanford.edu}
\affiliation{%
Leinweber Institute for Theoretical Physics, Stanford University, Stanford, California 94305, USA
}%
\author{Tatia Kiliptari}
\email{t.kiliptari@ufl.edu}
\affiliation{Department of Physics, University of Florida, Gainesville, FL 32611, USA}
\author{Dmitrii L. Maslov}
\email{maslov@phys.ufl.edu}
\affiliation{Department of Physics, University of Florida, Gainesville, FL 32611, USA}

\begin{abstract}
We develop a conserving relaxation-time approximation (cRTA) based on an explicit energy-resolved projection onto the full space of collision invariants. 
Our cRTA retains the energy dependence of the nonequilibrium quasiparticle distribution, allowing one to describe transport quantities sensitive to states near, but not exactly on, the Fermi surface (FS). We apply the method to several charge-transport problems  in both Galilean-invariant and non-Galilean-invariant Fermi liquids. In particular, the cRTA reproduces the low- and high-temperature limits of the dc conductivity of a non-Galilean-invariant Fermi liquid with disorder, the hydrodynamic and collisionless limits of the finite-wavevector longitudinal conductivity of a clean Galilean-invariant Fermi liquid, and the asymptotic scaling forms of the optical conductivity of a clean non-Galilean-invariant Fermi liquid beyond the semiclassical limit. For several observables, the agreement with exact solutions is quantitative at the percent level. These results demonstrate that the cRTA provides a simple and accurate framework for describing transport beyond the FS projection.
\end{abstract}
\maketitle

\paragraph{Introduction.}
The kinetic equation (KE) is a versatile tool for describing transport phenomena in both classical and quantum systems. However, its exact solutions are rare, even in linear response. The main difficulty arises from the non-locality of the collision integral in momentum space. Consequently,  the KE is an integro-differential one. Accordingly, a number of approximate methods for solving the KE were  developed in the past. 
A special category of these methods are relaxation-time approximations (RTAs), which replace the non-local collision integral for binary collisions  by a local one. The most common RTA with the collision integral 
$I_{\rm RTA}=-(f_\bk-n_\bk)/\tau$~\cite{ashcroft}, where $n_\bk$ is the equilibrium distribution, and $\tau$ is the phenomenological relaxation time,
does not enforce conservation laws. Conserving RTAs date back to the seminal paper by Bhatnagar, Gross, and Krook on mono-atomic classical gases~\cite{BKG:1954}. In condensed matter physics, partially and fully conserving RTAs for binary collision integrals were developed in Refs.~\cite{Callaway:1959,Mermin:1970,woelfle:1978,DeGennaro_1984, DeGennaro:1985,Molenkamp:1995,Conti:1999, atwal2002relaxation}.

Earlier conserving RTAs were formulated in terms of local thermodynamic variables and applied mainly to density response and hydrodynamics of electron gases with a parabolic dispersion \cite{atwal2002relaxation}. In this Letter, we formulate the conserving approximation as an explicit projection onto the hydrodynamic manifold, yielding a systematic scheme for describing transport in generic non-Galilean-invariant 
Fermi liquids. Our cRTA also differs from the previous methods in that it does not rely on a Fermi-surface  projection but operates with an energy-resolved projection onto the full space of collision invariants. This makes it applicable to transport quantities sensitive to the energy dependence of the nonequilibrium distribution near the Fermi surface (FS), e.g., the thermal conductivity of a generic Fermi liquid (FL)  and the charge conductivity of non-Galilean--invariant FLs. We show that our cRTA reproduces several exact and asymptotically exact results.

\paragraph{General construction.}\label{sec:Gen}

We consider a single-band system of electrons with arbitrary dispersion $\ve=\ve(\bk)$, measured from the Fermi energy. The interaction between electrons is parameterized by a phenomenological scattering time $\tee(\bk,T)$ which, in general, depends on the momentum $\bk$ and temperature $T$. Beyond the semiclassical limit, i.e., for arbitrary $\Omega/T$, where $\Omega$ is the frequency of the applied electric field, one also needs to include the dependence of $\tee$ on $\Omega$.

Electron-electron collisions conserve particle number, momentum (in the absence of umklapp scattering), and energy, such that a vector in the $d+2$ ``hydrodynamic'' space,
\bea
\vec\psi_{\bk}=\left(1,\ve,\bk\right),\eea
is conserved. Here and thereafter, $\vec a$ denotes a vector in this $d+2$ space, whereas $\mathbf{a}$ denotes a vector in real space. Our goal is to construct an algebraic form of the collision integral which  satisfies all the conservation laws. A deviation from equilibrium is parameterized  as 
\bea
\delta f_\bk&=&f_\bk-n_\bk =(-n_\bk') \chi_\bk,
\eea
where $n'_\bk\equiv \partial_\ve n(\ve)$. Introduce another deviation projected onto the hydrodynamic space:
\bea
\delta f^\mathrm{h}_\bk=(-n_\bk')\chi^\mathrm{h}_\bk,\quad
\chi^\mathrm{h}_\bk=\vec c\cdot \vec\psi_\bk
,\label{hydro}
\eea
where $\vec c$ is a $\bk$-independent $d+2$ vector.  

Consider the collision integral:
\bea
I_\mathrm{ee}[f_\bk]=-\frac{\delta f_\bk-\delta f^\mathrm{h}_\bk}{\tee}=-(-n_\bk')\frac{\chi_\bk-\chi_\bk^\mathrm{h}}{\tee}.\label{I}
\eea
If the vector $\vec c$ in Eq.~\eqref{hydro} is chosen as
\bea
\vec c=\hat{M}^{-1}\cdot 
\la\frac{\chi_\bk\vec\psi_\bk}{\tee}\ra,\label{vecc}
\eea 
with 
\bea
\langle X_\bk\rangle\equiv\int \frac{d^dk}{(2\pi)^d}
(-n'_\bk)X_\bk\label{average}
\eea
and 
\bea
M_{\alpha\beta}&=&M_{\beta\alpha}=
\la\frac{\psi_{\alpha\bk}\psi_{\beta\bk}}{\tee}\ra,\label{M}
\eea
then $I_\mathrm{ee}[f_\bk]$  in Eq.~\eqref{I} satisfies all the necessary conservation laws, i.e.,
\bea
\int \frac{d^dk}{(2\pi)^d}\,  \psi_{\alpha\bk}\, I_\mathrm{ee}[f_\bk]=0,\quad \alpha=1\dots d+2.\label{zero}
\eea
The above statement is proven 
as follows. Since $\int\frac{d^dk}{(2\pi)^d} \psi_{\alpha\bk} I_\mathrm{ee}[f_\bk]=\left\langle \psi_{\alpha\bk}(\chi_\bk^\mathrm{h}-\chi_\bk)/\tee\right\rangle$, Eq.~\eqref{zero} is equivalent to 
\bea
\la\frac{\psi_{\alpha\bk}\,\chi_\bk^\mathrm{h}}{\tee}\ra=
\la\frac{\psi_{\alpha\bk}\,\chi_\bk}{\tee}\ra.\label{task}
\eea
Substituting Eqs.~\eqref{vecc} and \eqref{M} into Eq.~\eqref{task} yields
\begin{centeq}
&\la\frac{\psi_{\alpha\bk}\,\chi_\bk^\mathrm{h}}{\tee}\ra=\la\frac{\psi_{\alpha\bk}}{\tee}\sum_\beta c_\beta \psi_{\beta\bk}\ra=\sum_\beta c_\beta M_{\alpha\beta}=\\
&\sum_{\beta,\gamma} (M^{-1})_{\beta\gamma}
\la\frac{\chi_\bk \psi_{\gamma\bk}}{\tee} \ra M_{\alpha\beta}=\la \frac{\chi_\bk\psi_{\alpha\bk}}{\tee}\ra,
\end{centeq}
as required. In what follows, we demonstrate that this method reproduces some known results.

\paragraph{\emph{dc} conductivity of a non-Galilean--invariant Fermi liquid with disorder.} By ``non-Galilean invariant'' we mean a system 
with a non-parabolic $\ve$ that may otherwise be an arbitrary function of $k\equiv |\bk|$, so that $\tee$ depends only on $\ve$ and $T$. Previous work has shown that
the electron-electron (ee) contribution to the conductivity depends on the ratio of the impurity  scattering time, $\ti$, and the \emph{quasiparticle} scattering time
due to the $ee$ interaction, 
$\tee(T)\propto T^{-2}$, with a $\ln T$ factor in two dimensions (2D)~\cite{kiliptari:2026}. At low temperatures ($\tee\gg\ti$), the conductivity exhibits a correction of order $T^4$
(again, with a $\ln T$ factor in 2D)  \cite{pal:2012b,Sharma:2021,Kovalev:2025,Kiliptari:2025}, and saturates at high $T$  ($\tee(T)\ll\ti$)~\cite{pal:2012b,Kiliptari:2025}. To mimic Coulomb interaction in 2D, 
we choose (in units where $\hbar=k_B=1$)
\bea\label{tee_form}
\frac{1}{\tee}=a\frac{(b\ve)^2+\pi^2 T^2}{\ef}\ln\frac{\Omega_{\rm p}}{T},\qquad T,\ve\ll \Omega_{\rm p},\label{tee}
\eea
where $\ve_\textrm{F}$ is the Fermi energy, $\Omega_{\rm p}=\vf\kappa$ is the characteristic plasma frequency, $v_\textrm{F}$ is the Fermi velocity,  $\kappa$ is the screening wavevector,
and $a$ and $b$ are numerical coefficients.
Although $b=1$ for the quasiparticle rate, we treat both $a$ and $b$ as fitting parameters.

We assume isotropic impurity scattering with an energy-independent 
$\ti$, and consider a 2D system with Dirac dispersion, as a particular example of a non-Galilean FL.  With $\mathbf{E}$ along the $x$-axis, the linearized semiclassical KE reads
\bea
ev_{x}E (-n_\bk')=I_\mathrm{ee}[f_\bk]-\frac{f_\bk-\bar f_\bk}{\ti},\label{FLdis}
\eea
where $\bv=\partial_\bk \ve$ and $\bar O_\bk$ is the average of $O_\bk$ over the directions of $\bk$. 
The impurity part of the collision integral is exact for point-like disorder. 

Since the non-equilibrium distribution is proportional to $v_x E$, $\langle \chi_\bk/\tee\rangle=\langle\chi_\bk \ve/\tee\rangle=\langle \chi_k k_y/\tee \rangle=0$. From Eq.~\eqref{vecc}, $c_1=c_\ve=c_{y}=0$, and also $\bar \chi_\bk=0$. Substituting Eq.~\eqref{I} with $c_1=c_\ve=0$ into Eq.~\eqref{FLdis} yields 
\bea
\chi_\bk=\frac{\ttot}{\tee}c_x k_x-\ttot ev_x E,\label{chi}
\eea
where $\ttot = \tee\ti/(\tee+\ti)$, whereas Eq.~\eqref{vecc} gives,
\bea
c_x=\left(\hat{M}^{-1}\right)_{xx} \la \frac{\chi_\bk k_x}{\tee}\ra.\label{cx}
\eea
For an isotropic system, the matrix $\hat M$ is block-diagonal  in the scalar $(1,\ve)$ and vector ($\bk$) subspaces, with $\left(\hat{M}^{-1}\right)_{xx}=1/\langle k_x^2/\tee\rangle$.  Equations~\eqref{chi} and~\eqref{cx} then yield
\bea
c_x=-\frac{eE\la\frac{\ttot}{\tee} v_x k_x\ra}{\la\frac{1}{\tee+\ti}k_x^2\ra}.
\eea
The conductivity is thus 
\bea
\sigma=2e^2 \ti
\left(\left\langle\frac{\tee}{\tee+\ti}v_x^2\right\rangle+\ti\frac{\left\langle \frac{1}{\tee+\ti} k_xv_x
\right\rangle^2}{\left\langle\frac{1}{\tee+\ti} k_x^2\right\rangle}\right).\label{sigmaeei}
\eea

For a parabolic dispersion with band mass $m$ $(k_x=m v_x$), $\tee$ drops out from the result,
 and the conductivity is controlled entirely by impurities: $\sigma=\sigma_\mathrm{i}=2e^2\ti\la v_x^2\ra$. 
The same holds after FS projection, i.e., for $v=\vf$ and $k=\kf$ (Fermi momentum). 
A finite \textit{ee} contribution thus requires expanding $k$ and $v$ near the FS.

For $\tee\gg \ti$, expanding 
Eq.~\eqref{sigmaeei} in $\ti/\tee$ yields a correction $\delta\sigma = \sigma-\sigma_{i0}$ to the residual conductivity, $\sigma_{\mathrm{i}0}=\sigma_{\rm i}\vert_{T=0}$,
\bea
\delta\sigma=-2e^2\ti^2 \frac{\left\langle\frac{v_x^2}{\tee}\right\rangle\left\langle\frac{k_x^2}{\tee}\right\rangle-\left\langle\frac{v_x k_x}{\tee}\right\rangle^2}{\left\langle\frac{k_x^2}{\tee}\right\rangle}.\label{lowT}
\eea
 For Dirac spectrum, $v_F={\rm constant}$, $k=\kf(1+\ve/\vf)$, and the density of states $\nu(\ve)=\nf(1+\ve/\ef)$, where $\nf=\ef/2\pi \vf^2$.  
Evaluating the averages~\cite{SM_dc}, we find
\bea
\delta\sigma=
-\frac{\pi^3 a}{30}(7b^2+5)
e^2\ti^2\frac{T^4}{\ef^2}\ln\frac{\Omega_{\rm p}}{T}.\label{dclowT}
\eea
For $\tee\ll\ti$, $\tee$ drops out from the leading-order result:
\bea
\sigma&=& 2e^2\ti \frac{\langle k_x v_x\rangle^2}{\langle k_x^2\rangle},\label{highT}
\eea
in agreement with Refs.~\cite{pal:2012b,Kiliptari:2025}.

\begin{figure}[h!]
    \centering
\includegraphics[width=\linewidth]{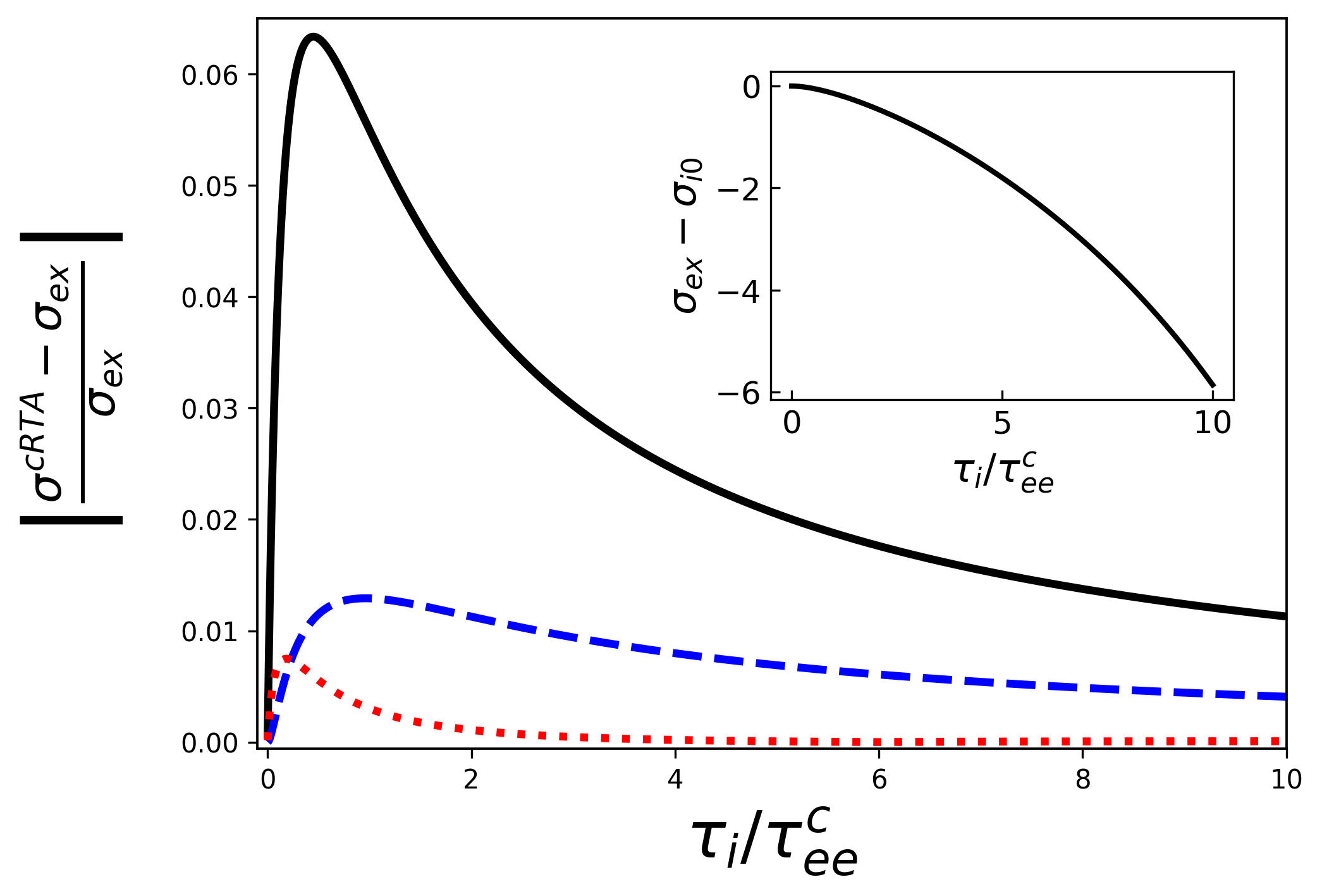} 
    \caption{Main panel: Relative deviation between the cRTA and exact  results,
    Eqs.~\eqref{sigmaeei} and~\eqref{scalingF2}, respectively.  
    Solid (black), 
    dashed (blue), and 
    dotted (red)
    lines correspond to $b=0, 1, 0.732$ in Eq.~\eqref{tee}, respectively. Inset:  the $T$-dependent part of the exact solution in units of $e^2$, for $\Omega_{\rm p}^2 \ti / \ef = 100$. \label{fig:bench}
}
\end{figure}

The exact conductivity for the same model, was obtained in~\cite{kiliptari:2026}:
\bea
\sigma_{\rm ex}&=&\sigma_{\mathrm{i}0}
+\frac{e^2T^2\ti}{\ef}
\mathcal{F}\!\left(\frac{\tau^{\rm C}_{\rm ee}}{\ti}\right),\label{scalingF2}
\eea
where $\tee^{\rm C}=4\ve_{\rm F}/\pi T^2\ln(\Omega_{\rm p}/T)$ for the Coulomb interaction, and
the function $\mathcal{F}(z)$ is expressed in terms of the associated Legendre functions and obeys $\mathcal F(x\gg 1)=-4\pi/15x$ and $\mathcal F(x\ll 1)=-\pi/6+\pi x/8$~\cite{SM_exact}. The $T$ dependence of the second term in Eq.~\eqref{scalingF2} is shown in the inset to Fig.~\ref{fig:bench}.

To benchmark Eq.~\eqref{sigmaeei}, we compare it to the low-$T$ limit of the exact result~\cite{kiliptari:2026},
\bea
\delta\sigma_{\rm ex}=\sigma_{\rm ex}-\sigma_{{\rm i}0}=- \frac{\pi^2}{15} e^2\ti^2\frac{T^4}{\ef^2}\ln\frac{\Omega_{\rm p}}{T},\label{exlowT}
\eea
which has the same structure as Eq.~\eqref{scalingF2}.  We choose $a=2/\pi(7b^2+5)$; then $\delta\sigma=\delta\sigma_{\textrm{ex}}$ at low temperatures.
The main panel of Fig.~\ref{fig:bench}
shows the relative deviation between the cRTA result (Eq.~\ref{sigmaeei}) and exact result (Eq.~\ref{scalingF2}) for three choices of $b$:
$b=0$ (energy-independent quasiparticle rate); $b=1$ (actual quasiparticle rate); and $b\approx0.732$, chosen as the best-fit value over the entire temperature range. The 
relative errors are $\lesssim6\%$, $\sim1\%$, and below $1\%$, respectively.

\paragraph{Longitudinal conductivity of a clean Galilean-invariant Fermi liquid at finite wavevector.} We now consider the longitudinal conductivity $\sigma_L$ of a clean, $d$-dimensional, Galilean--invariant FL subject to an external electric field with frequency $\Omega\ll\ef$ and wavevector $q\ll\kf$, while the ratio $\Omega/\vf q$ is arbitrary. Unlike the previous example, this problem probes both the particle-number and momentum sectors of the hydrodynamic space. The conductivity remains finite after FS projection, so that the average Eq.~\eqref{average} is taken only over the FS. We also neglect the energy dependence of the \emph{ee} scattering time,  which affects only numerical coefficients: $\tee=\tee(T)\equiv1/\gamma$.

With $\bE \: ||\: \bq \: || \: \hat{{\mathbf x}}$, the KE reads
\bea
(\gamma-i\Omega+iq v_x)\chi_\bk=-e E v_x+\gamma(c_1+c_\ve \ve_\bk+c_x k_x).
\eea
After FS projection, $\hat M$ becomes diagonal:
$\hat{M}={\rm diag}(M_{11},M_{\ve\ve},M_{xx})$. Furthermore, $c_\ve=0$, because all terms in $\chi_\bk$, except the one proportional to $\ve_\bk$, vanish upon integration over $\ve_\bk$. The remaining components of Eq.~\eqref{vecc} form a $2\times2$ system for $c_1$ and $c_x$, which yields~\cite{SM_finiteq}
\begin{equation}
\sigma_{\rm L}(q,\Omega)
=
\frac{ne^2}{m}
\frac{
(1-\gamma{\cal I}_0){\cal I}_2+\gamma{\cal I}_1^2
}{
(1-\gamma{\cal I}_0)(1/d-\gamma{\cal I}_2)
-\gamma^2{\cal I}_1^2
},
\label{sigmaL}
\end{equation}
where $n$ is the electron number density,
\bea
{\cal I}_n\equiv
\left\langle
\frac{\mu^n}{\gamma-i\Omega+iq\vf\mu}
\right\rangle_{\rm FS},
\eea
$\mu=\mathbf{k}\cdot \mathbf{q}/ k q$, and $\la O_\bk\ra_{\rm FS}$ denotes the angular average taken after the FS projection. 

In the hydrodynamic limit ($\Omega,\vf q\ll\gamma$), Eq.~\eqref{sigmaL} reduces to 
\begin{equation}
\sigma_L(q,\Omega)
=
\frac{ne^2 }{m}
\,
\frac{i\Omega}
{\Omega^2-c_{\rm s}^2q^2+i\nu_{\rm L}\Omega q^2},
\label{sigma-hydro}
\end{equation}
where $c_{\rm s}=\vf/\sqrt{d}$ is the sound velocity and 
\begin{equation}
\nu_{\rm L}=
\frac{2(d-1)}{d+2}
\frac{c_{\rm s}^2}{\gamma}. 
\label{nuL}
\end{equation}
The exact hydrodynamic result is of the same form as Eq.~\eqref{sigma-hydro}, with $\nu_{\rm L}$ replaced by the longitudinal kinematic viscosity~\cite{Landau:H,Gurzhi:1968,Conti:1999},
\bea
\nu^{\rm h}_{\rm L}=2\frac{d-1}{d} \eta+\zeta.
\label{nuH}
\eea
Equations~\eqref{nuL} and~\eqref{nuH} imply that 
$\zeta=0$, which is correct for a monoatomic gas with parabolic dispersion~\cite{physkin}, and that the shear viscosity $\eta=0$ in $d=1$~\cite{Conti:1999}. 
With $\gamma\propto T^2$ (up to 
$\ln T$ in 2D), Eq.~\eqref{nuL} reproduces the scaling 
$\eta\propto1/T^2$.

In the collisionless limit ($\Omega,\vf q\gg\gamma$), Eq.~\eqref{sigmaL} yields 
\bea
\R\!\sigma_L(q,\Omega)
=
-\frac{ne^2}{m}\!\frac{d \gamma}{\vf^2 q^2}
\left[
{\cal J}_2'\left(z\right)+{\cal J}_1^2\left(z\right)+d{\cal J}_2^2(z)
\right],\label{nocoll}
\eea
where $z=\Omega/\vf q>1$ and ${\cal J}_n=\la\mu^n/(z-\mu)\ra_{\rm FS}$. In the viscoelastic limit of the collisionless regime ($\Omega\gg \vf q\gg\gamma$)~\cite{Conti:1999}, Eq.~\eqref{nocoll}
is simplified to
\bea
\R\sigma_{\rm L}(q,\Omega)
=
\frac{ne^2}{m}
\frac{2(d-1)}{d(d+2)}
\frac{v_F^2q^2\gamma}{\Omega^4},
\label{viscoel}
\eea
which agrees with the  microsocopic result $\R\sigma_{\rm L}(q,\Omega)=g_d e^2 \kf^{d-2} \vf^2 q^2 T^2/\Omega^4$,
where $g_d$ is the Coulomb coupling constant in $d$ dimensions~\cite{Sharma:2024}, for an appropriate choice of the coefficient $a$ in  $\gamma=a T^2$.

 \paragraph{Optical conductivity of a clean Fermi liquid beyond the semiclassical limit.}
Both previous examples apply in 
the semiclassical limit: $\Omega\ll T$. To extend our analysis for an arbitrary ratio $\Omega/T$, the KE needs to be modified in two ways. First, the semiclassical driving term  in Eq.~\eqref{FLdis} is replaced by $e(\bv\cdot\bE)\Delta(\Omega,\ve_\bk)$, where 
\bea
\Delta(\Omega,\ve_\bk)=\frac{1}{\Omega}\left[n\left(\ve_\bk-\frac{\Omega}{2}\right)-n\left(\ve_\bk+
\frac{\Omega}{2}\right)\right].\label{Delta}
\eea
This result is obtained 
by Wigner-transforming the left-hand side of the Dyson equation for the lesser non-equilibrium Green's function~\cite{SM_Keldysh}, 
\begin{equation*}
\Big[i(\partial_{t_1}+\partial_{t_2})
-\varepsilon_{\mathbf{k}+e\mathbf{A}(t_1)}
+\varepsilon_{\mathbf{k}+e\mathbf{A}(t_2)}\Big]G^<_\bk(t_1,t_2).
\end{equation*}
Accordingly, we re-define $\delta f_\bk$ as $\delta f_\bk=\Delta(\Omega,\ve_\bk)\chi_\bk$ and the average in Eq.~\eqref{average} as 
$\langle X\rangle_\Omega\equiv\int 
d^dk
\Delta(\Omega,\ve_\bk)X/(2\pi)^d.$
The second modification accounts for the effect of the electric field on the Keldysh collision integral, $I_{\rm k}$, which contains Moyal products of the self-energies $\Sigma^{>,<}_\bk$ and Green's functions $G^{>,<}_\bk$.  Let $X_0$ be the equilibrium part of $X$, $\delta X$ its deviation, and
\bea
\hat S_{\omega,\Omega}\left[X\right]\equiv \frac{1}{2}\left[X(\omega+\Omega/2)+X(\omega-\Omega/2)\right].
\eea
Evaluating the Moyal products for a harmonic perturbation, we obtain~\cite{SM_Keldysh}
\begin{centeq}
    I_{\rm K}=-\int \frac{d\omega}{2\pi 
}&\{\delta\Sigma^<_\bk\hat S_{\omega,\Omega}\left[G^>_{0\bk}\right]+\delta G^>_\bk\hat S_{\omega,\Omega}\left[\Sigma^<_{0\bk}\right]-\\
&\delta\Sigma^>_\bk\hat S_{\omega,\Omega}\left[G^<_{0\bk}\right]-\delta G^<_\bk\hat S_{\omega,\Omega}\left[\Sigma^>_{0\bk}\right]\}.\label{IK}
\end{centeq}
Within the cRTA, the frequency shift by $\Omega/2$ can be modeled by replacing $1/\tee$ with
\bea
\frac{1}{\tee^\Omega}=\frac{1}{2}\left[\frac{1}{\tee(\ve_\bk+\Omega/2,T)}+\frac{1}{\tee(\ve_\bk-\Omega/2,T)}\right].\label{Omega}
\eea
With these modifications, the cRTA kinetic equation becomes
\bea
-i\Omega 
\chi_\bk+e(\bE\cdot\bv)=-\frac{\chi_\bk-\chi_\bk^{\rm h}}{\tee^\Omega},
\eea
and the general scheme is the same as before with the replacements $\la\dots\ra\to\la\dots\ra_\Omega$ and $\tee\rightarrow \tee^\Omega$.

As a check, we apply Eq.~\eqref{Omega} to the simplest case of a clean, three-dimensional (3D) FL with an $ee$ interaction that does not conserve momentum. Accordingly, ${\mathbf c}=0$. Since $c_1=c_\ve=0$ in linear response as well, $\chi_\bk^{\rm h}=0$ and we obtain $(i,j=x,y,z)$
\bea
\sigma_{ij}=2e^2\la\frac{v_i v_j}{1/\tee^\Omega-i\Omega}\ra_\Omega.
\eea
Choosing $1/\tee(\ve,T)=A(\ve
^2+\pi^2 T^2)$, we obtain in the collisionless regime ($1/\tee(0,T)\ll\Omega\ll\ef$)
\bea
\R\sigma_{ij}=\frac{2e^2}{\Omega^2}\la\frac{v_i v_j}{\tee^\Omega}\ra_\Omega\propto \frac{\Omega^2+4\pi^2 T^2}{\Omega^2},
\eea
which is the canonical Gurzhi scaling~\cite{gurzhi:1959}.

To probe Eq.~\eqref{Omega} further, we turn to a clean, isotropic, but non-Galilean-invariant FL. To be specific, we  consider 3D Weyl fermions at finite Fermi energy and assume that
$1/\tee=a_0(\ve^2+\pi^2 T^2)/\ef$.
The real part of the optical conductivity consists of a $
\delta(\Omega)$ piece, which implies that the \emph{dc} conductivity is infinite due to momentum conservation, and a regular piece, $\sigma_\mathrm{ee}$, which is finite for
$\Omega>0$. The result for the conductivity can be obtained from Eq.~\eqref{sigmaeei} by replacing $\ti\to i/\Omega$, $\tee\to\tee^\Omega$, and $\la\dots\ra\to\la\dots\ra_\Omega$. Subtracting off the the reactive high-frequency tail, we obtain (with $\mathbf{E}$ along the $z$-axis)
\bea
\see= i\frac{e^2}{\Omega}\left[-\left\langle \frac{v_z^2}{1-i\alpha}\right\rangle_\Omega+\frac{\left\langle \frac{k_z v_z}{1-i\alpha}\right\rangle_\Omega^2}{\left\langle\frac{k_z^2}{1-i\alpha}\right\rangle_\Omega}\right],\label{seeRTA}
\eea 
where $\alpha=\Omega\tee^\Omega$. As in the \emph{dc} case, the integrands of $\la\dots\ra_\Omega$ need to be expanded around the Fermi energy to obtain a finite result.

In the the collisionless regime, $\Omega\gg 1/\tee(\ve=0,T)$,
Eq.~\eqref{seeRTA} yields
\bea
\R\see=\frac{e^2
a_0}{180\pi} 
\frac{(\Omega^2+4\pi^2 T^2)(\Omega^2+6\pi^2 T^2)}{\Omega^2\vf\ef}.\label{highreq}
\eea
This equation reproduces the general features of the result of a microscopic calculation in Ref.~\cite{Sharma:2021}, up to the numerical coefficient in the second bracket: the exact coefficient is $8\pi^2/3$ rather than $6\pi^2$. The difference comes about because, in a microscopic calculation, an $\mathcal{O}(\ve_\bk^2)$ term in the expansion near $\ef$ also acquires a shift by $\pm\Omega/2$, whereas the cRTA misses this effect. At the same time, the first bracket in Eq.~\eqref{highreq} ensures that the first bosonic Matsubara frequency rule is satisfied, i.e., that $\R\sigma_\mathrm{ee}\vert_{\Omega=\pm 2\pi i T}=0$~\cite{maslov:2012}.

In the hydrodynamic limit, $0<\Omega\ll 1/\tee(\ve=0,T)$,
the conductivity saturates at the $T$-independent value
\bea
\R\sigma_\mathrm{ee}=
\frac{1}{6\pi^2 
a_0}\left(1-\frac{\pi^2}{12}\right) e^2 \kf+\mathcal{O}(T^2).
\eea\label{lowfreqex}
With $a_0=(\pi^3/56)(1-\pi^2/12)\kappa / k_F$, this coincides with  the exact solution of Ref.~\cite{kiliptari:2026}, extended to the 3D case.

In conclusion, we presented a simple but versatile conserving relaxation-time approximation (cRTA), which accounts for all required conservation laws and is applicable to situations where  FS projection is insufficient. We anticipate that this version of the cRTA can be applied to several other situations, including non-linear response~\cite{Sayak:unpub}. 

We thank S. Raghu, E. Tulipman, Y.-M. Wu and J. Yu for useful conversations. SB is supported in part by the US Department of Energy, Office of Basic Energy Sciences, Division
of Materials Sciences and Engineering, under Contract
No. DE-AC02-76SF00515. TK and DLM aknowledge support from the National Science Foundation via Grant No. DMR-2224000.

The data that support the findings of this article are 
available at Ref.~\cite{kiliptari2026datacrtavsexact}.

\bibliography{references}
\clearpage

\setcounter{page}{0}
\setcounter{equation}{0}
\setcounter{section}{0}

\renewcommand{\thepage}{S\arabic{page}}
\renewcommand{\theequation}{S\arabic{equation}}
\renewcommand{\thesection}{S\arabic{section}}

\setcounter{secnumdepth}{3}

\begin{widetext}
\begin{center}
{\Large
    Conserving relaxation-time approximation
    for electron-electron collisions\\
    \bigskip Sayak Bhattacharjee, Tatia Kiliptari, and Dmitrii L. Maslov\\
    \bigskip Supplemental Material}
\end{center}

\section{\emph{dc} conductivity of a non-Galilean--invariant Fermi liquid with disorder}
\label{sec:dc}
Below we show some steps in analyzing Eq.~\eqref{sigmaeei} on the MT.
\subsection{Parabolic spectrum}
Substituting $k_x=m v_x$ into Eq.~\eqref{sigmaeei}, we find
\bea
\sigma&=&2e^2 
\ti\left[\left\langle\frac{\tee}{\tee+\ti}v_x^2\right\rangle+\ti\frac{\left\langle \frac{1}{\tee+\ti} v^2_x
\right\rangle^2}{\left\langle\frac{1}{\tee+\ti} v_x^2\right\rangle}\right]\nn\\
&=&2e^2 
\ti\left[\left\langle\frac{\tee}{\tee+\ti}v_x^2\right\rangle+\ti\left\langle \frac{1}{\tee+\ti} v^2_x
\right\rangle\right]=2e^2\ti\langle v_x^2\rangle.
\eea
Hence $\tee$ drops out, as it should.

\subsection{Low-$T$ regime: $\tee\gg \ti$} Expanding Eq.~\eqref{sigmaeei} to lowest order in $\ti/\tee$, we obtain
\bea
\sigma=
\sigma_{\rm i}+\delta\sigma
\eea
where $\sigma_{\rm i}=2e^2 \ti\langle v_x^2\rangle$, 
\bea
\delta\sigma=-2e^2\ti^2 \frac{\left\langle\frac{v_x^2}{\tee}\right\rangle\left\langle\frac{k_x^2}{\tee}\right\rangle-\left\langle\frac{v_x k_x}{\tee}\right\rangle^2}{\left\langle\frac{k_x^2}{\tee}\right\rangle},\label{pertdc}
\eea
and we remind the reader that 
\bea
\la X \ra\equiv \int \frac{d^dk}{(2\pi)^d} (-n'_\bk)X.\label{averageSM}
\eea

If the integrals in $\la\dots\ra$ are confined to the FS, the two terms in the numerator of $\delta\sigma$ cancel each other. To get a finite result, one needs to expand near the FS, taking into account that $\tee$ is even in $\ve_\bk$. For 2D Dirac fermions, $\vf={\rm consant}$, $k=\kf(1+\ve/\ef)$, and $\nu(\ve)=\nf(1+\ve/\ef)$.  With  $\ve_\bk\equiv \ve$ and $W(\ve)\equiv (-n')/\tee$, we expand the numerator to leading order in $\ve/\ef$ to obtain
\bea
&&\left\langle\frac{v_x^2}{\tee}\right\rangle\left\langle\frac{k_x^2}{\tee}\right\rangle-\left\langle\frac{v_x k_x}{\tee}\right\rangle^2=
\frac{1}{4}\nf^2\vf^2 \kf^2\left\{\int d\ve W(\ve) \left(1+\frac{\ve}{\ef}\right) \int d\ve' W(\ve') \left(1+\frac{\ve'}{\ef}\right)^3\right.\nn\\
&&\left.-\left[\int d\ve W(\ve) \left(1+\frac{\ve}{\ef}\right)^2
\right]^2\right\}\approx \frac{1}{4}\nf^2 \vf^2 \kf^2 \left[ 3\int d\ve W(\ve)\int d\ve' W(\ve') \frac{\ve'^2}{\ef^2}\right.\nn\\
&&\left.-2\int d\ve W(\ve) \int d\ve' W(\ve') (\ve'/\ef)^2\right]
=\frac{1}{4}\nf^2 \vf^2 \kf^2 \int d\ve W(\ve) \int d\ve' W(\ve') \frac{\ve'^2}{\ef^2}.
\label{expand}
\eea
To the same order, $k_x^2$ in the denominator of Eq.~\eqref{pertdc} can be projected onto the FS.
Hence
\bea
\delta\sigma=-
e^2\nf \vf^2\ti^2 \int d\ve W(\ve) \frac{\ve^2}{\ef^2}.\label{deltas}
\eea
For 
\bea
\frac{1}{\tee}=a\frac{(b\ve)^2+\pi^2 T^2}{\ef}\ln\frac{\Omega_{\rm p}}{T},
\eea
Eq.~\eqref{deltas} 
leads to Eq.~\eqref{dclowT} of the MT.

\subsection{High $T$ regime: $\tee\ll\ti$} 
For $\tee\ll \ti$, Eq.~\eqref{sigmaeei} of MT is  simplified to
\bea
\sigma&=&2e^2 
\ti\left[\left\langle\frac{\tee}{\tee+\ti}v_x^2\right\rangle+\ti\frac{\left\langle \frac{1}{\tee+\ti} k_xv_x
\right\rangle^2}{\left\langle\frac{1}{\tee+\ti} k_x^2\right\rangle}\right]\nn\\
&\approx & 2e^2 
\ti\left[\frac{\langle k_x v_x\rangle^2}{\langle k_x^2\rangle}+\left\langle\frac{\tee}{\ti}v_x^2\right \rangle\right]\approx 2e^2\ti \frac{\langle k_x v_x\rangle^2}{\langle k_x^2\rangle},
\eea
which is exactly the same as obtained by a spectral decomposition of the $ee$ collision integral~\cite{pal:2012b}.
\section{Results of an exact solution of the kinetic equation}
\label{sec:exact}

For the reader's convenience, we present below the exact result for the \emph{dc} conductivity of a 2D FL with Coulomb interaction, a Dirac dispersion and disorder~\cite{kiliptari:2026}. 
The KE equation is solved by reducing it to the second order inhomogeneous equation of the Legendre type. With 
\bea
\lambda\equiv\tee^{\rm C}(T)/\ti,\quad \beta \equiv \sqrt{1 + \lambda
},
\eea
the Green's function of this equation is given by
\begin{align}
\mathcal{R}_\beta(s,s') = \mathcal{C_\beta} \,
\begin{cases}
P^{-\beta}_1(s) P^{\beta}_1(s'), & s > s',\\
P^{-\beta}_1(s') P^{\beta}_1(s), & s < s',
\end{cases}
\label{Gf2}
\end{align}
where $\mathcal{C_\beta}=1/2\beta(\beta^2-1)$ and \begin{align}
&P^{\pm\beta}_1(s)= (s\mp\beta) \left( \frac{1+s}{1-s} \right)^{\pm\beta/2}. 
\label{psib}
\end{align}
are the associated Legendre functions. In terms of the quantities defined above,
the scaling function $\cal{F}$ in Eq.~\eqref{scalingF2} of the main text (MT) is given by
\bea
\mathcal{F}(\lambda)\equiv
2\pi\int_0^1 ds s\int_{-1}^1 ds' s'\,\frac{
\sqrt{1-s'^2}
}{\sqrt{1 - s^2}}\mathcal{R}_\beta(s,s').\label{functionF}
\eea
\section{Longitudinal conductivity of a clean Galilean-invariant Fermi liquid at finite wavevector}
\label{sec:finiteq}
\subsection{General solution}

Consider a Galilean-invariant FL in $d$ dimensions, subject to a spatially non-uniform and oscillatory electric field. The electric field $\bE$ is parallel to $\bq$, and both are chosen along the $x$-axis. At finite $q$, the longitudinal conductivity  is finite, and the leading contribution to the conductivity comes from electrons on the FS. Therefore, $\la\dots\ra$ is reduced to averaging over the FS. Since the energy dependence of the $ee$ scattering rate affects only the numerical coefficients, we assume it to be energy independent. Neglecting also the logarithmic factor in 2D, we take
\bea
\frac{1}{\tee}\equiv \gamma=\frac{gT^2}{\ef}.
\eea

A finite-\(q\) kinetic equation in the cRTA reads 
\begin{equation}
L_{\bk}\chi_{\bk}
=
-eE v_x+\gamma(c_1+c_\ve\ve_{\bk}+c_x k_x),
\qquad
L_{\bk}\equiv \gamma-i\Omega+iqv_x .
\label{eq:KE-finite-q-const-gamma-full}
\end{equation}
or
\bea
\chi_{\bk}
=
\frac{-eE k_x/m+\gamma(c_1+c_x k_x)}{L_{\bk}}
+
c_\ve\,\frac{\gamma\ve_{\bk}}{L_{\bk}} .
\eea
Since $\gamma={\rm constant}$, it drops out from Eq.~\eqref{vecc} of the MT. We re-define the average (Eq.~\ref{average}) projecting it onto the Fermi surface, and obtain 
\bea
\la X\ra=\nf\int \frac{d\mathcal{O}_\bk}{\mathcal{O}_d}  X_\bk\Big\vert_{k=k_F}\equiv \nu_F \la X_\bk\ra_{\rm FS},
\eea
where $d\mathcal{O}_\bk$ and $\mathcal{O}_d$ are the element of and full solid angle in $d$ dimensions, respectively.
After this projection, the matrix $\hat M$ in Eq.~\eqref{M} of the MT becomes diagonal:
\bea
\hat M
=\nf\,{\rm diag}(1,\ve_F^2,k_F^2/d).
\eea
It then follows that
\bea
c_\ve=\frac{1}{M_{\ve\ve}}\gamma\la \ve_\bk\chi_\bk\ra=0,
\eea
because $\chi_\bk$ is even in $\ve_\bk$.

The coefficients $c_1$ and $c_x$ are found from the $2\times 2$ system 
\bea
c_1&=& \frac{1}{M_{11}} \la \chi_\bk\ra\nn\\
c_x&=& \frac{1}{M_{xx}}  \la \chi_\bk k_x\ra\label{c1cx}
\eea
Substituting
\[
\chi_{\bk}
=
\frac{-eEv_x+\gamma(c_1+c_xk_x)}{L_{\bk}},
\]
into Eq.~\eqref{c1cx} and using \(v_x=k_x/m\), gives
\begin{align}
c_1
&=
-eE\vf{\cal I}_1
+\gamma c_1{\cal I}_0
+\gamma c_x \kf {\cal I}_1,
\\
\frac{1}{d}c_x 
&=
-\frac{eE}{m}{\cal I}_2
+\frac{\gamma}{\kf} c_1{\cal I}_1
+\gamma c_x{\cal I}_2,
\end{align}
where
\bea
{\cal I}_n\equiv 
\left\langle
\frac{\mu^n}{\gamma-i\Omega+iq\vf\mu}
\right\rangle_{\rm FS}\label{Igen}
\eea
and $\mu$ is the cosine of the angle between $\bk$ and $\bq$. Solving for $c_1$ and $c_x$, and calculating the current $j_x=-2e\la v_x\chi_\bk\ra$, we obtain the longitudinal conductivity as
\begin{equation}
\sigma_{\rm L}(q,\Omega)
=
\frac{e^2 n}{m}
\frac{
(1-\gamma{\cal I}_0){\cal I}_2+\gamma{\cal I}_1^2
}{
(1-\gamma{\cal I}_0)(1/d-\gamma{\cal I}_2)
-\gamma^2{\cal I}_1^2
},
\label{sigmaL_SM}
\end{equation}
where we used that $2\nu_F \vf^2/d=n/m$ for a parabolic dispersion.
The integrals entering Eq.~\eqref{sigmaL_SM} can be calculated exactly. 
It is convenient to introduce
\bea
s\equiv \frac{\Omega+i\gamma}{\vf q}.\label{sdef}
\eea
Then
$\gamma-i\Omega+iq\vf\mu
=
-i \vf q(s-\mu)$
and therefore
\bea
{\cal I}_n
=
\frac{i 
}{\vf q}\,{\cal J}_n(s),
\qquad
{\cal J}_n(s)
\equiv
\left\langle
\frac{\mu^n}{s-\mu}
\right\rangle_{\rm FS} .
\eea
The first three angular integrals are related by
\[
{\cal J}_1=s{\cal J}_0-1,
\qquad
{\cal J}_2=s^2{\cal J}_0-s,
\]
Thus it is enough to compute
\({\cal J}_0\).

In \(d=2\),
\[
{\cal J}_0(s)
=
\int_0^{2\pi}\frac{d\theta}{2\pi}
\frac{1}{s-\cos\theta}
=
\frac{1}{\sqrt{s^2-1}},
\]
where the branch is chosen such that
\bea
{\cal J}_0^{(2)}(s)\to \frac{1}{s}
\qquad
{\rm for}\;|s|\to\infty.\label{branch}
\eea
Hence
\bea
{\cal I}_0
=
\frac{i
}{\vf q}
\frac{1}{\sqrt{s^2-1}},\;{\cal I}_1^{(2)}
=
\frac{i 
}{\vf q}
\left(
\frac{s}{\sqrt{s^2-1}}-1
\right),\;{\cal I}_2^{(2)}
=
\frac{i }{\vf q}
\left(
\frac{s^2}{\sqrt{s^2-1}}-s
\right).\label{I2d}
\eea

In \(d=3\),
\bea
{\cal J}_0^{(3)}(s)
=
\frac12\int_{-1}^{1}\frac{d\mu}{s-\mu}
=
\frac12\ln\frac{s+1}{s-1},
\eea
again with the branch chosen according to the condition~\eqref{branch}. Therefore,
\bea
{\cal I}_0
=
\frac{i}{2\vf q}
\ln\frac{s+1}{s-1},\;{\cal I}_1=\frac{i 
}
{\vf q}
\left[
\frac{s}{2}\ln\frac{s+1}{s-1}-1
\right],\;{\cal I}_2
=
\frac{i }{\vf q}
\left[
\frac{s^2}{2}\ln\frac{s+1}{s-1}-s
\right].\label{I3d}
\eea

\subsection{Hydrodynamic limit}
We now consider the hydrodynamic limit
\[
\Omega\ll\gamma,\qquad v_Fq\ll\gamma,
\]
but do not assume any hierarchy between \(\Omega\) and \(v_Fq\).  Although one could expand the exact results from Eqs.~\eqref{I2d} and~\eqref{I3d} in $\Omega/\gamma$ and $\vf q/\gamma$, it is more instructive to expand first the integrand of Eq.~\eqref{Igen} and then perform the angular averaging.

Denoting
\bea
L_{\bk}=\gamma+a_{\bk},
\qquad
a_{\bk}\equiv -i\Omega+iq\vf\mu,
\eea
we have
\bea
\frac{1}{L_{\bk}}
=
\frac{1}{\gamma}
\left[
1-\frac{a_{\bk}}{\gamma}
+\frac{a_{\bk}^2}{\gamma^2}
-\cdots
\right].
\eea
Keeping terms needed for the denominator in Eq.~\eqref{sigmaL}, we find
\begin{subequations}
\begin{align}
1-\gamma{\cal I}_0
&=
\frac{\langle a_{\bk}\rangle_{\rm FS}}{\gamma}
-\frac{\langle a_{\bk}^2\rangle_{\rm FS}}{\gamma^2}
+O(\gamma^{-3})
,
\\
1/d-\gamma{\cal I}_2
&=
\frac{
\langle 
\mu^2 a_{\bk}\rangle_{\rm FS}}{\gamma}
-\frac{\langle 
\mu^2 a_{\bk}^2\rangle_{\rm FS}}{\gamma^2}
+O(\gamma^{-3}),
\\
\gamma{\cal I}_1
&=
-
\frac{\langle 
\mu a_{\bk}\rangle_{\rm FS}}{\gamma}
-\frac{\langle 
\mu a_{\bk}^2\rangle_{\rm FS}}{\gamma^2}
+O(\gamma^{-3}).
\end{align}
\end{subequations}
Noting that  
$\langle\mu^4\rangle_{\rm FS}=3/d(d+2)$
and recalling that the sound velocity of a Fermi gas is given by
\bea
c_{\rm s}=\vf/\sqrt{d},
\eea
we obtain
\begin{align}
\langle a_{\bk}\rangle_{\rm FS}
&=-i\Omega,
\\
\langle 
\mu^2 a_{\bk}\rangle_{\rm FS}
&=-i\Omega\, 
/d,
\\
\langle 
\mu a_{\bk}\rangle_{\rm FS}
&=i\frac{
q \vf
}
{d
},
\\
-\langle a_{\bk}^2\rangle_{\rm FS}
&=\Omega^2+c_{\rm s}^2 q^2
,
\\
-\langle 
\mu^2a_{\bk}^2\rangle_{\rm FS}
&=\Omega^2\frac{
1}{d}+\frac{3c_{\rm s}^2q^2
}{
d+2},
\\
\langle 
\mu a_{\bk}^2\rangle_{\rm FS}
&=2\Omega \frac{\vf q\
}{d
}.
\end{align}

Substituting the expansions above into the denominator of Eq.~\eqref{sigmaL}, we find 
\bea
\Delta\equiv\Delta
=
(1-\gamma{\cal I}_0)(1/d-\gamma{\cal I}_2)
-\gamma^2{\cal I}_1^2\approx\frac{1}{d\gamma^2} \left[c_{\rm s}^2q^2-\Omega^2+\frac{i\Omega}{\gamma}\left(-2\Omega^2+\frac{6}{d+2} c_{\rm s}^2q^2\right)\right].\label{den}
\eea
Although the  $i\Omega/\gamma$ term is subleading, it must be kept in the denominator to account for damping of the sound mode.

The expand the numerator of Eq.~\eqref{sigmaL} we also need
\begin{align}
{\cal I}_2
&=
\left[
\frac{1}
{d\gamma}
-\frac{\langle \mu^2a_{\bk}\rangle_{\rm FS}}{\gamma^2}
+\frac{\langle \mu^2a_{\bk}^2\rangle_{\rm FS}}{\gamma^3}.
+O(\gamma^{-4})
\right].
\end{align}
Hence
\bea
{\cal N}\equiv
(1-\gamma{\cal I}_0){\cal I}_2+\gamma{\cal I}_1^2
\approx-\frac{\nf^2 \kf^2}{d} \frac{i\Omega}{\gamma^2}\left(1+\frac{2i\Omega}{\gamma}\right).
\eea
The $i\Omega/\gamma$ correction is irrelevant near the sound pole. Neglecting it,
\bea
{\cal N}\approx-\frac{\nf^2 \kf^2}{d} \frac{i\Omega}{\gamma^2}\label{num}
\eea

Substituting Eq.~\eqref{den} and Eq.~\eqref{num} into Eq.~\eqref{sigmaL} of the MT, we obtain
\bea
\sigma_{\rm L}(q,\Omega)=\frac{ne^2}{m}\frac{i\Omega
}{
\Omega^2-c_{\rm s}^2q^2 +i\frac{\Omega}{\gamma}\left(2\Omega^2 -\frac{6}{
d+2
} 
c_{\rm s}^2 q^2
\right)}.
\eea
Near the sound pole $|\Omega|\approx c_{\rm s}q$, one can replace $|\Omega|$ by $c_{\rm s}q$ in the damping term in the denominator. Thus the conductivity is reduced to
\begin{equation}
\sigma_{\rm L}(q,\Omega)
=
\frac{e^2 n}{m}
\,
\frac{i\Omega}
{\Omega^2-c_{\rm s}^2q^2+i\nu_L\Omega q^2},\qquad \nu_{\rm L}=
\frac{2(d-1)}{d+2}
\frac{c_{\rm s}^2}{\gamma},
\label{sigma-hydro-SM}
\end{equation}
which is Eq.~\eqref{sigma-hydro} of the MT.

We now obtain the same result purely within hydrodynamics. We write the linearized Navier-Stokes equation as~\cite{Landau:H}
\bea
mn\partial_t {\mathbf u}=-\boldsymbol{\nabla} P-e{\bf E}+mn\eta\nabla^2 {\mathbf u}+mn\left(\zeta+\eta\left(1-\frac{2}{d}\right)\right)\boldsymbol{\nabla}(\boldsymbol{\nabla}\cdot{\mathbf u}),
\eea
where $\mathbf{u}$ is the hydrodynamic velocity, $n$ is the uniform number density,  $\eta$ and $\zeta$ are the kinematic shear and bulk viscosities, correspondingly, and $P$ is the pressure. For ${\bf E}=E_0\hat x e^{iqx-\Omega t}$,
\bea
-imn\Omega u_x=-\partial_x \delta P-eE_0
+mn \nu^{\rm H}_{\rm L}\partial^2_x u_x,\label{NS}\eea
where
\bea
\nu^{\rm H}_{\rm L}=\zeta+2\frac{d-1}{d} \eta\label{nuHapp}.
\eea
The isoentropic relation
\bea
\left(\frac{\partial P}{\partial n}\right)_{S}=mc_{\rm s}^2 
\eea
remains valid for slowly varying perturbations. Then 
$\delta P= mc_{\rm s}^2 \delta n$
or, in the momentum space, 
$iq \delta P=c_{\rm s}^2 iq \delta n$.
Eliminating $\delta n$ from the continuity equation
\bea
-i\Omega \delta n+iq n u_x=0,
\eea
we find
\bea
iq\delta P=i\frac{m n c_{\rm s}^2 q^2}{\Omega} u_x
\eea
Substituting this back into Eq.~\eqref{NS}, solving for $u_x$, and finding the current as $j_x=-en u_x$, we obtain
\begin{equation}
\sigma^H_L(q,\Omega)
=
\frac{e^2 n}{m}
\,
\frac{i\Omega}
{\Omega^2-c_{\rm s}^2q^2+i\nu^H_L\Omega q^2}
\label{sigma-hydro-NS}
\end{equation}
which is of the same form as Eq.~\eqref{sigma-hydro}. 
Comparing the viscosities in Eqs.~\eqref{nuHapp} and ~\eqref{sigma-hydro-SM}, we see that they agree if $\zeta=0$, as it should be the case for a monoatomic gas with parabolic dispersion \cite{physkin}.

\subsection
{Collisionless limit}
We now turn to the collisionless limit, when  \(\Omega\gg \gamma\) and \(\vf q\gg\gamma\), while the ratio $\Omega/\vf q$ is arbitrary.
Expanding Eq.~\eqref{sigmaL} formally to first order in \(\gamma\), we obtain 
\bea
\sigma_{\rm L}(q,\Omega)=
d\frac{ne^2}{m}\left[
{\cal I}_2+\gamma\left({\cal I}_1^2+d{\cal I}_2^2\right)
\right]
+O(\gamma^2).
\eea
Now we expand the first term in the small imaginary part of \(s\), as defined by Eq.~\eqref{sdef}:
\[
{\cal I}_2
=
\frac{i}{\vf q}{\cal J}_2\!\left(\frac{\Omega}{\vf q}+i\frac{\gamma}{\vf q}\right)
=
\frac{i}{\vf q}{\cal J}_2\left(\frac{\Omega}{\vf q}\right)
-\frac{\gamma}{\vf^2q^2}{\cal J}_2'\left(\frac{\Omega}{\vf q}\right)
+O(\gamma^2).
\]
In the \(O(\gamma)\) terms, one may set \(s=\Omega/\vf q)\). 
Collecting everything together,
\bea
\sigma_{\rm L}(q,\Omega)
=d\frac{ne^2}{m}\left\{
\frac{i 
}{\vf q}{\cal J}_2\left(\frac{\Omega}{\vf q}\right)
-
\frac{
1}{\vf^2q^2}
\left[
{\cal J}_2'\left(\frac{\Omega}{\vf q}\right)+{\cal J}_1^2\left(\frac{\Omega}{\vf q}\right)+d{\cal J}_2^2\left(\frac{\Omega}{\vf q}\right)
\right]
\right\}
+O(\gamma^2).
\eea
For \(\Omega>\vf q\), the functions \({\cal J}_n(\Omega/\vf q)\) are real, while the zeroth-order term is purely imaginary. Therefore
\bea
\R\sigma_{\rm L}(q,\Omega)
=
-d\frac{e^2n}{m}
\frac{\gamma}{\vf^2q^2}\left[
{\cal J}_2'\left(\frac{\Omega}{\vf q}\right)+{\cal J}_1^2\left(\frac{\Omega}{\vf q}\right)+d{\cal J}_2^2\left(\frac{\Omega}{\vf q}\right)
\right]
,
\qquad \Omega>\vf q,\label{collfree}
\eea
which is Eq.~\eqref{nocoll} of the MT.

Despite an overall minus sign in Eq.~\eqref{collfree}, 
$\R\sigma_{\rm L}$ is non-negative, as it should be. 
To see that this is the case, we define
\bea
f(\mu)\equiv \frac{\mu}{z-\mu},\quad z=\Omega/\vf q,
\eea
so that
\bea
{\cal J}_1(z)=\langle f\rangle_{\rm FS},
\qquad
{\cal J}_2(z)=\langle \mu f\rangle_{\rm FS},\qquad {\cal J}_2'(z)
=
-\left\langle
\frac{\mu^2}{(z-\mu)^2}
\right\rangle_{\rm FS}
=
-\langle f^2\rangle_{\rm FS}.
\eea
After the FS projection, the conserved hydrodynamic subspace is reduced to $(1,k_x)=(1,\kf \mu)$ or just to $(1,\mu)$.
The projection of \(f\) onto the  subspace spanned by \(1\) and \(\mu\) is
\[
Pf
=
\langle f\rangle_{\rm FS}
+
d\langle \mu f\rangle_{\rm FS}\,\mu ,
\]
because
\[
\langle 1\rangle_{\rm FS}=1,
\qquad
\langle \mu\rangle_{\rm FS}=0,
\qquad
\langle \mu^2\rangle_{\rm FS}=\frac{1}{d}.
\]
Hence
\[
\langle (f-Pf)^2\rangle_{\rm FS}
=
\langle f^2\rangle_{\rm FS}
-
\langle f\rangle_{\rm FS}^2
-
d\langle \mu f\rangle_{\rm FS}^2 .
\]
and therefore
\[
{\cal J}_2'(z)+{\cal J}_1^2(z)+d{\cal J}_2^2(z)
=
-\langle (f-Pf)^2\rangle_{\rm FS}\le 0,
\]
so that $\R\sigma_{\rm L}(q,\Omega)\geq 0$.

Finally, we consider  the viscoelastic limit of the collisionless regime \cite{Conti:1999},
when $\gamma\ll\vf q\ll \Omega$ and thus $z\gg 1$.
The required large-\(z\) expansions are
\begin{align}
{\cal J}_1(z)
&=
\frac{1}{d z^2}
+
\frac{3}{d(d+2)z^4}
+
O(z^{-6}),
\\
{\cal J}_2(z)
&=
\frac{1}{d z}
+
\frac{3}{d(d+2)z^3}
+
O(z^{-5}),
\\
{\cal J}_2'(z)
&=
-\frac{1}{d z^2}
-
\frac{9}{d(d+2)z^4}
+
O(z^{-6}).
\end{align}
Therefore
\begin{align}
{\cal J}_2'(z)+{\cal J}_1^2(z)+d{\cal J}_2^2(z)
&=
-\frac{2(d-1)}{d^2(d+2)}
\frac{1}{z^4}
+
O(z^{-5}).
\end{align}

Substituting these expansions into Eq.~\eqref{collfree}
gives
\begin{equation}
\R\sigma_{\rm L}(q,\Omega)
=
\frac{ne^2}{m}
\frac{2(d-1)}{d(d+2)}
\frac{\vf^2q^2\gamma}{\Omega^4},
\qquad
\gamma\ll \vf q\ll \Omega,
\label{eq:Re-sigma-viscoelastic-limit}
\end{equation}
which is Eq.~\eqref{viscoel} of the MT.
\section{kinetic equation beyond the semiclassical limit}
\label{sec:Keldysh}
\subsection{Electric-field driving}
To derive the electric-field driving term beyond the semiclassical limit, i.e., for arbitrary $\Omega/T$ ratio, we temporarily omit the collision integral.
In the presence of a time-dependent and uniform electric field, the left-hand side of the Dyson equation for the  lesser non-equilibrium Green's function is
\begin{equation}
\hat L G^<_\bk(t_1,t_2)\equiv \Big[i(\partial_{t_1}+\partial_{t_2})
-\varepsilon_{\mathbf{k}+e\mathbf{A}(t_1)}
+\varepsilon_{\mathbf{k}+e\mathbf{A}(t_2)}\Big]G^<(t_1,t_2,\bk).
\label{Dyson}
\end{equation}
To linear order in the field
\begin{equation}
\varepsilon_{\mathbf{k}+e\mathbf{A}(t_{1,2})}-\ve_\bk=
\,e\,\bv\!\cdot\!\mathbf{A}(t_{1,2})=\frac{\,e\,\bv_{\mathbf{k}}\!\cdot\!\mathbf{E}}{i\Omega}e^{-i\Omega t_{1,2}}.
\end{equation}
To the same order, $G^<_\bk(t_1,t_2)$ multiplying $\bE$ can be replaced by the equilibrium function, $G^<_{0\bk}(t_1-t_2)$, which depends only on the time difference $t_1-t_2$.

A Wigner transform of a function $X(t_1,t_2)$ is defined as
 \bea
 X(\omega,t)\equiv \int d\tau e^{i\omega\tau} X(t+\tau/2,t-\tau/2),
 \eea
where $t=(t_1+t_2)/2$ and $\tau=t_1-t_2$.
Wigner transforming Eq.~\eqref{Dyson} with the help of 
\bea
\int d\tau e^{i\omega\tau} \left[e^{-i\Omega(t+\tau/2)}-e^{-i\Omega(t-\tau/2)}\right]\int\frac{d\omega'}{2\pi} G^<_{0\bk}(\omega')e^{-i\omega'\tau}
=e^{-i\Omega t}\left[G^<_{0\bk}(\omega-\Omega/2)-G^<_{0\bk}(\omega+\Omega/2)\right],
\eea
we obtain
\bea
\hat L G^<_\bk
(\omega,t)=i\partial_t G^<_\bk(\omega,t)-\frac{\,e\,\bv\!\cdot\!\mathbf{E}\,e^{-\Omega t}}{i\Omega}\left[G^<_0(\omega-\Omega/2,\bk)-G^<_0(\omega+\Omega/2,\bk)\right].\label{WL}
\eea
In the quasiparticle approximation,  the distribution function is defined as \cite{physkin,rammer:1986}
\bea
G^<_\bk(\omega, t)=2\pi i\delta(\omega-\ve_\bk) f_\bk(t),\;G^<_{0\bk}(\omega, t)=2\pi i\delta(\omega-\ve_\bk) n(\ve_\bk).
\eea
Substituting these definitions into Eq.~\eqref{WL} and integrating over $\omega$, we obtain the driving term in Eq.~\eqref{Delta} of the MT.

\subsection{Collision integral}

The right-hand side of the Dyson equation for $G^<_\bk $ reads
\bea
\tilde I(\bk,\omega,t)=\frac{1}{2}\left[\Sigma^<_\bk \star G^>_\bk-\Sigma^>_\bk\star G^<_\bk+G^>\star \Sigma^<_\bk-G^<_\bk\star\Sigma^>_\bk\right](\omega,t), \label{tildeI}
\eea
where a Moyal product is defined as
\bea
C(\omega,t)=(A\star B)(\omega,t)=\int d\tau e^{i\omega(t_1-t_2)}\int dt_3 A(t_1,t_3) B(t_3,t_2)
\eea
with $t=(t_1+t_2)/2$ and $\tau=t_1-t_2$. Shifting the integration variable as $t_3=t+s$,  
\bea
C(\omega,t)=\int d\tau e^{i\omega\tau}  \int dsA(t+\tau/2,t+s) B(t+s,t-\tau/2).
\eea
In A, the ``center time'' and ``difference time" are $(1/2)(t+\tau/2+t+s)=t+\tau/4+s/2$ and $\tau/2-s$, respectively. In $B$, the center time and difference time are $t-\tau/4+s/2$ and $\tau/2+s$, respectively. Expressing $A$ and $B$
via their Wigner transforms, we find
\bea
C(\omega,t)&=& 
\int d\tau \int ds \int \frac{d\omega_1}{2\pi}  \frac{d\omega_2 }{2\pi}  e^{i\left[\omega-(1/2))(\omega_1+\omega_2)\right]\tau} e^{i(\omega_1-\omega_2)s}  A(\omega_1, t+\tau/4+s/2) B(\omega_2,t+s/2-\tau/4).\label{WM}
\eea 

Suppose that the dependence on the center-time is harmonic: $A(\omega,t)=e^{-i\Omega_A t}A_{\Omega_A}(\omega)$ and $B(\omega,t)=e^{-i\Omega_B t}B_{\Omega_B}(\omega)$ with $\Omega_A,\Omega_B=0,\Omega$. Substituting these forms into \eqref{WM}, we obtain
\bea
C(t,\omega)
&=&e^{-i(\Omega_A+\Omega_B)t}
\int d\tau e^{i\omega\tau}\int ds \int \frac{d\omega_1}{2\pi}  \frac{d\omega_2 }{2\pi}  e^{i\left[\omega-(1/2))(\omega_1+\omega_2)-\frac{\Omega_A-\Omega_B}{4}\right]\tau} e^{i\left(\omega_1-\omega_2-\frac{\Omega_A+\Omega_B}{2}\right)s}  A_{\Omega_A}(\omega_1) B_{\Omega_B}(\omega_2)\nn\\
&=&e^{-i(\Omega_A+\Omega_B)t}\int d\omega_1 \int d\omega_2
\delta\left(\omega_1-\omega_2-\frac{\Omega_A-\Omega_B}{4}\right)\delta\left(\omega_1-\omega_2-\frac{\Omega_A+\Omega_B}{2}\right)A_{\Omega_A}(\omega_1) B_{\Omega_B}(\omega_2)\nn\\
&=&e^{-i(\Omega_A+\Omega_B)t} A_{\Omega_A}(\omega+\Omega_B/2) B_{\Omega_B}(\omega-\Omega_A/2)\label{WMosc}
\eea 

In what follows, we will be interested in linear response, when $A=\mathcal{A}_0+e^{-i\Omega t} \delta A$ and $B=B_0+e^{-i\Omega t} \delta B$ with time-independent $A_0$ and $B_0$.  To linear order in $\delta A$ and $\delta B$,
\bea
C(\omega,t)&=&\left(\mathcal{A}_0+\delta A e^{-i\Omega t}\right)\star \left(B_0+\delta B e^{-i\Omega t}\right)\nn\\
&=& \mathcal{A}_0 B_0 +\left(\delta A e^{-i\Omega t}\right)\star B_0 +\mathcal{A}_0\star \left(\delta B e^{-i\Omega t}\right)\nn\\
&=& \mathcal{A}_0B_0+\left[\delta A(\omega) B_0(\omega-\Omega/2)  +\mathcal{A}_0(\omega+\Omega/2) \delta B(\omega)\right]e^{-i\Omega t}.
\label{harmonic}
\eea
Linearizing the Moyal products of $\Sigma^{<,>}_\bk$ and $G^{<,>}_\bk$ in Eq.~\eqref{tildeI} around their corresponding equilibrium values, applying Eq.~\eqref{harmonic} to each term, and integrating over $\omega$, we obtain the collision integral \eqref{IK} of the MT.

\end{widetext}
\end{document}